# Failure Analysis in Transition: An Industry Survey of Challenges, Priorities, and Standardization Needs in Advanced Packaging and Heterogeneous Integration

Himanandhan Reddy Kottur*, Nusra Akter Takia*, Mahamudul Hassan Fuad*, Istiaq Firoz Shiam*, Matthew Walsh[ⱡ], Navid Asadizanjani*
*Department of Electrical and Computer Engineering, University of Florida, Gainesville, FL, USA
[ⱡ] Florida Semiconductor Engine, Kissimmee, FL, USA
Corresponding authors: h.kottur@ufl.edu, nasadi@ufl.edu

***Abstract*—**Failure analysis is being reshaped by heterogeneous integration, chiplet-based architectures, hybrid bonding, backside technologies, & increasingly buried package structures. To examine how practitioners view this transition, an anonymous survey was distributed across a broad set of organizations involved in semiconductor design, packaging, systems, tools, & failure analysis. The survey collected approximately one hundred responses & probed organizational background, supported product domains, future priorities in failure analysis, critical bottlenecks, sample preparation challenges, emerging architecture specific pain points, & perceived needs for workflow acceleration & data standardization. The results show that heterogeneous integration, chiplet, and three-dimensional products dominate the respondent base at 69%, while package & heterogeneous integration failure analysis received the highest importance rating at 7.92 out of 10. Hybrid bonding emerged as the most difficult new architecture to analyze at 54%, higher-resolution non-destructive imaging ranked as the most important future accelerator at 8.18 out of 10, and 83% of respondents supported formalized data standardization frameworks. The complete survey data are provided in Appendix A (Table II) to improve transparency & support future benchmarking.

***Keywords*—***failure analysis, heterogeneous integration, advanced packaging, hybrid bonding, chiplets, non-destructive imaging, sample preparation, standardization, artificial intelligence.*

## I. INTRODUCTION

Failure analysis has traditionally been associated with device level localization, physical deprocessing, materials characterization, & root cause isolation in comparatively accessible semiconductor structures. That model remains important, but the structure of modern products has changed. Fig 1. represents the evolution of semiconductor failure analysis from conventional device-centric methodologies toward multi-abstraction-level analysis required for advanced heterogeneous integration technologies. Advanced packaging, fan out technologies, chiplets, three-dimensional stacking, backside power delivery, & hybrid bonding have redistributed failure mechanisms across interfaces, buried interconnects, package materials, thermal pathways, & assembly induced stresses. As a result, failure analysis now operates across multiple abstraction levels rather than within only the transistor or interconnect domain.

This transition creates two related pressures. i) physical: regions of interest are harder to access, geometries are more fragile, & destructive preparation can distort the defect being investigated. ii) informational: meaningful diagnosis increasingly depends on combining electrical, physical, thermal, imaging, process, & package context into one coherent interpretation. The field therefore requires not only better instrumentation, but also stronger workflow integration & better data handoff among tools & teams.

To study how practitioners perceive this change, a survey was conducted:

- Across participants from fabless companies, OSAT & packaging organizations, systems & OEM companies, EDA & tool vendors, failure analysis laboratories, universities, & other institutions.
- The survey was focused on the practical barriers that shape present day failure analysis, especially in advanced packaging & heterogeneous integration.

The goal of this paper is not simply to list responses, but to interpret them as evidence of a broader shift in the center of gravity of failure analysis. The complete survey results are presented in Appendix A (Table I) for full transparency.

## II. SURVEY SCOPE & RESPONDENT PROFILE

The survey collected approximately one hundred anonymous responses from participants spanning multiple parts of the semiconductor ecosystem. Organizational representation included:

a) OSAT & packaging organizations at 22%
b) Fabless companies at 19%
c) Failure analysis laboratories at 16%
d) Systems & OEM organizations at 10%
e) EDA & tool vendors at 10%
f) And an additional 23% categorized as other

The comments within the other category indicate participation from academic institutions, modeling & simulation groups, prototyping & testing organizations, & specialized material suppliers, which broadens the perspective beyond only high-volume manufacturing. Fig. 2. represents the percentage distribution of key focus areas in modern semiconductor failure analysis, emphasizing packaging, fabless ecosystems, EDA/tool vendors, and AI-assisted analysis methodologies.

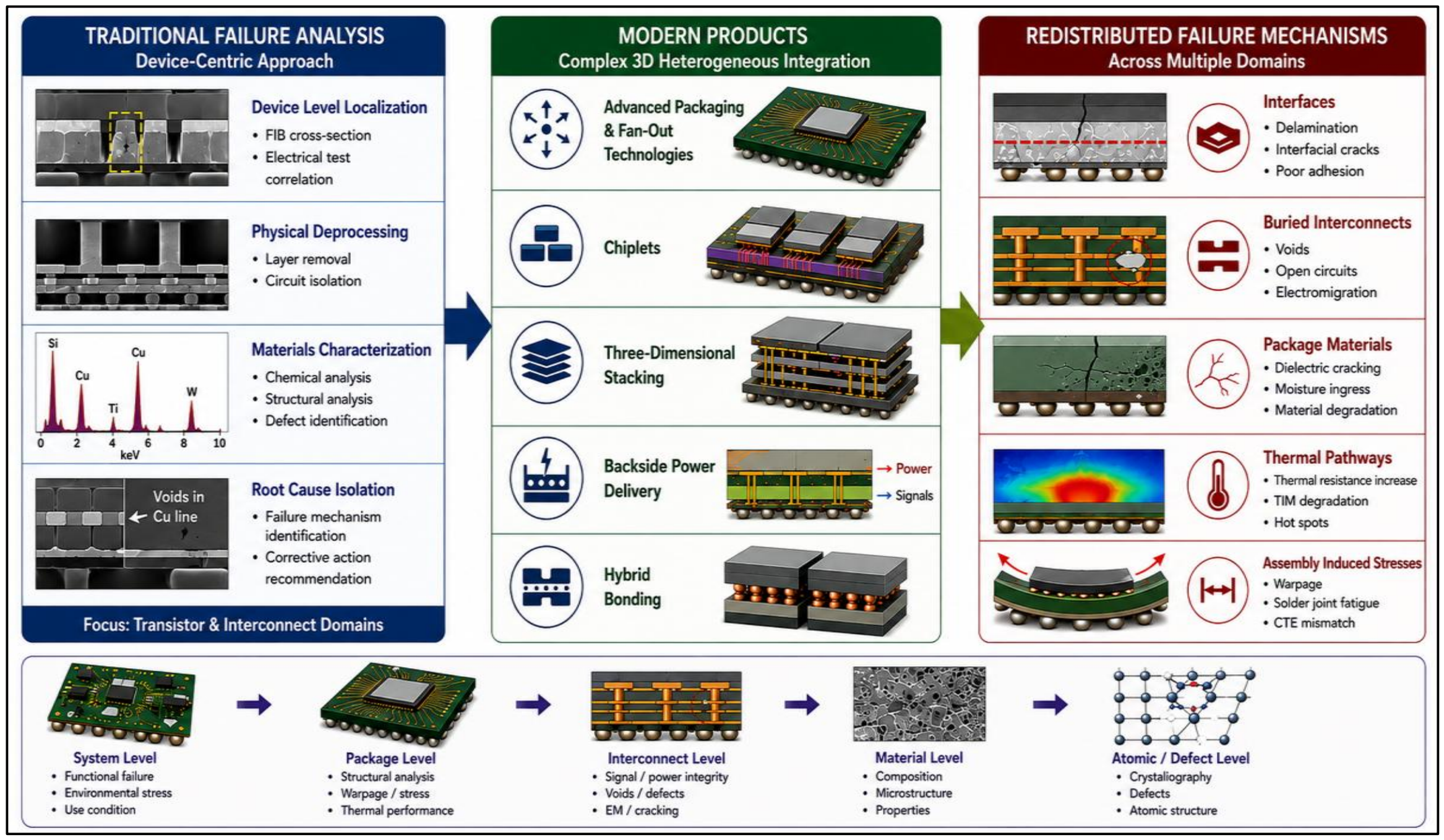

Fig. 1. Evolution of Failure Analysis in Advanced Heterogeneous Semiconductor Packaging Architecture.

The supported product domains reveal where the strongest technical concentration currently lies. Heterogeneous integration, chiplet, & three-dimensional products accounted for 69% of responses, the highest share in the survey. Photonics & co packaged optics followed at 41%, while logic at advanced nodes represented 35%, power products 24%, MEMS & sensors 20%, memory 16%, & other product categories 14%. This distribution is significant because the dominant product base already consists of architectures where failures can emerge from interfaces, packaging complexity, thermomechanical coupling, & cross-die interactions rather than from isolated transistor-level defects.

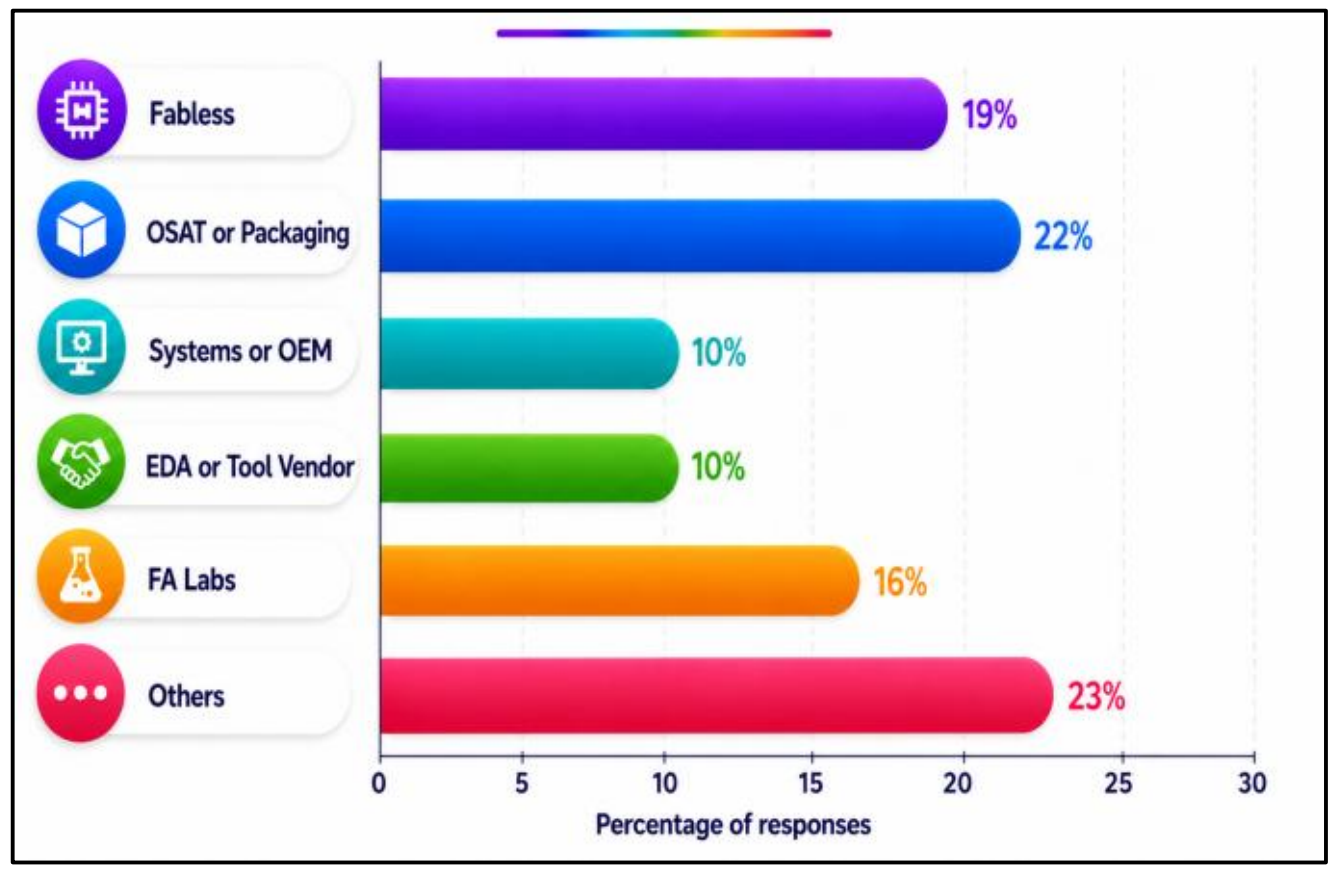

Fig. 2. Distribution of survey participants by organization category

Table I summarizes key quantitative highlights from the survey. These metrics are discussed in detail in the sections that follow; readers are encouraged to cross-reference Table II in Appendix A for the full result set.

**TABLE I.** ***Key Quantitative Survey Highlights***

| Metric | Value |
|---|---|
| Package & HI FA importance rating | **7.92/10** |
| System-level FA importance rating | **7.56/10** |
| Device-level fault isolation rating | **7.51/10** |
| Device-level physical analysis rating | **7.03/10** |
| Hybrid bonding: hardest architecture | **5.4/10** |
| BSPDN: backside probing as top challenge | **7.5/10** |
| NDI imaging: highest future accelerator | **8.18/10** |
| Support for standardization frameworks | **8.3/10** |

## III. IMPORTANCE OF FAILURE ANALYSIS AREAS

Participants were asked to rate the importance of several failure analysis domains for the coming years on a ten-point scale (see Table I). The results are instructive: they do not imply that device-level methods are losing value; rather, they indicate that package- and system-coupled problems now demand equal strategic attention alongside traditional device-level work.

Package & heterogeneous integration failure analysis received the highest average score at 7.92 out of 10. System-level failure analysis followed at 7.56, device-level fault isolation at 7.51, & device-level physical analysis at 7.03. This ranking is especially notable because package & heterogeneous integration analysis exceeds both traditional device level catagories. In practical terms, this reflects the diagnostic burden imposed by die to wafer integration, fan out packaging, large substrates, buried interfaces, cross domain signal pathways, & thermomechanical interactions that are not captured by conventional device centric workflows. The survey therefore supports the view that failure analysis is expanding upward from transistor level structures toward interfaces, assemblies, & field relevant system behavior.

## IV. CURRENT BOTTLENECKS IN WORKFLOW EXECUTION

The survey also examined the most critical organizational challenges. Although individual ratings varied by category, the strongest pressures clustered around package level defects, preparation related access problems, & the difficulty of integrating standardized data with artificial intelligence-based interpretation. These results should be read as a coupled problem rather than as isolated complaints. In advanced packaging, a defect may only become interpretable if the sample is opened correctly, the buried region is imaged with sufficient resolution, & the resulting data can be correlated with electrical behavior, process history, & design intent.

The open-ended responses reinforce this interpretation. Participants asked for the ability to directly connect failure mechanisms to functional behavior, for stronger modeling & verification capability, & for workflows that can handle increasing product complexity without disproportionate cost & latency. Several comments explicitly stated that traditional failure analysis is becoming too slow, too costly, & in some cases insufficient for determining the true root cause in highly integrated systems. This concern is consistent with the quantitative responses & suggests that workflow scalability is now a major technical issue.

## V. SAMPLE PREPARATION & IMAGING LIMITATIONS

Sample preparation remains one of the most consequential bottlenecks identified in the survey. The listed concerns include preparing ultra-thin lamella without introducing artifacts, achieving accurate delayering in advanced BEOL & low k structures, accessing regions of interest such as BSPDNs, TSVs, TGVs, & hybrid bonded interfaces, & consistently preparing large or non-standard geometries affected by bow or warpage. These are not minor procedural inconveniences. In many advanced packages, the act of exposing the structure can change fracture paths, redistribute stress, smear interfaces, or remove evidence needed for reliable root cause identification.

A related survey question asked which imaging or analysis methods most need improvement. Non-destructive analysis methods were selected far more often than destructive analysis methods, with:

- 72% of respondents favoring greater improvement in non-destructive approaches versus
- 28% for destructive approaches.

This result is especially important because it demonstrates a strong industry demand for deeper observability before irreversible deprocessing occurs. The open-ended responses in this section repeatedly request: i) better resolution, ii) greater penetration depth, iii) faster throughput, iv) improved reconstruction, v) multimodal imaging, vi) the ability to localize faults within true three-dimensional volumes without disturbing the assembly [1-2] vii) nano-void detection and high-resolution 3D X-ray, & viii) And better understanding of how artificial intelligence can be used under limited data & trustworthiness constraints. At the same time, destructive analysis requests remain technically demanding, including cryogenic SEM for fragile materials, low dose TEM, sub angstrom metrology, & ultra-thin lamella preparation. Together, these findings suggest that future failure analysis will rely on a tighter interplay between minimally invasive screening & carefully targeted destructive confirmation.

## VI. EMERGING ARCHITECTURES & THEIR PAIN POINTS

The survey provides particularly clear insight into architecture specific difficulty. When respondents were asked which, new architecture is hardest to analyze today, hybrid bonding received 54% of responses, well above gate all around devices at 24%, backside power delivery networks at 14%, & other categories at 8%. This is a strong signal that hybrid bonding has become a central challenge in the current failure analysis landscape.

The hybrid bonding specific responses identify three dominant technical obstacles: (i) detecting voids, gaps, & misalignment at the bonding interface (ii) achieving cross section preparation without inducing artifacts and& (iii) managing warped wafers or dies & the resulting impact on bonding yield. These results reveal why hybrid bonding is so challenging: diagnosis requires simultaneous control of interface imaging, sample preparation fidelity & package level mechanics. The failure signature may be local, but its origin can be coupled to global warpage, alignment error, oxide condition, copper topography & process induced stress [3].

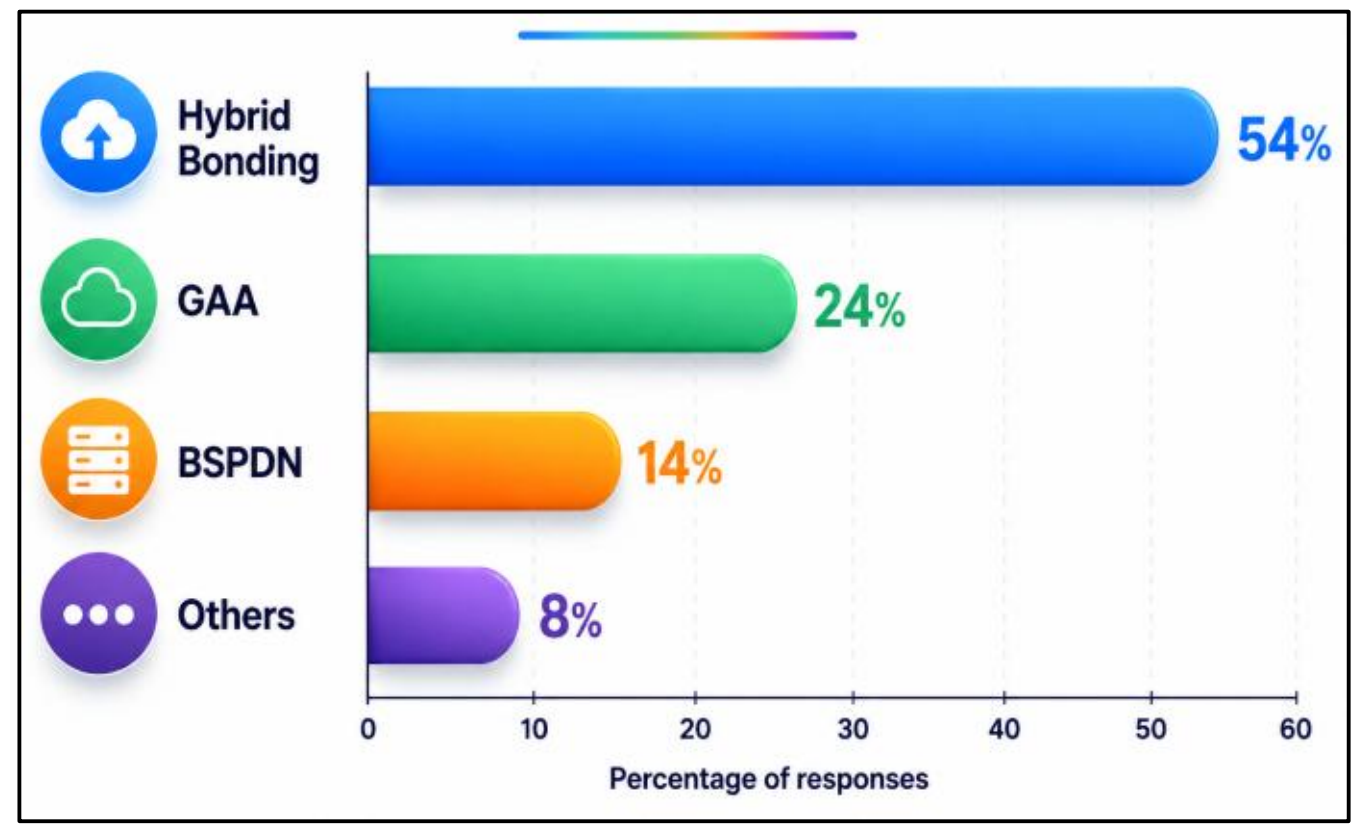

Fig. 3. Most challenging emerging device & packaging architectures for failure analysis

Gate all around technologies present a different but related challenge profile. For GAA analysis, 56% of respondents identified structural damage during electrical probing or nanoprobing as the dominant issue, while 44% cited delayering without losing device integrity. This distribution shows that access itself is becoming a failure analysis risk. In highly scaled structures, the act of preparing or contacting the device can alter the evidence under investigation. Similarly, for backside power delivery networks, 75% of respondents identified electrical probing or nanoprobing from the backside as the most difficult issue, while 25% pointed to locating & isolating defects in buried power rails. These responses confirm that the dominant challenge in BSPDN analysis is not only localization, but also practical measurement access through an increasingly concealed architecture [4]. Fig. 3. illustrates the percentage distribution of emerging semiconductor technology priorities among respondents. Hybrid bonding received the highest interest, followed by gate-all-around (GAA) devices and backside power delivery networks (BSPDN), reflecting the industry's transition toward advanced heterogeneous integration and next-generation device architectures.

Comments on overall three-dimensional heterogeneous integration extend this picture. Participants cited heat dissipation, electromagnetic compatibility, active embedded components, die to die fault isolation, CTE mismatch, hybrid bonding quality, power delivery, signal integrity, & interactions across multiple dies & abstraction levels. Such responses show that failure analysis in 3DHI is fundamentally a multiscale problem. The root cause may no longer map neatly onto one process module or one structural layer, which raises the need for better integration between package design, process knowledge, characterization, & reliability interpretation.

## VII. WHAT WOULD MOST ACCELERATE FUTURE FAILURE ANALYSIS

Among the proposed areas of improvement, higher resolution non-destructive imaging for buried structures & advanced packaging received the strongest average score at 8.18 out of 10. Artificial intelligence & machine learning enabled defect detection & classification followed at 7.27 out of 10. Integration of standardized data from multimodal sources such as electrical failure analysis, physical failure analysis, X ray, & data fusion rather than from isolated tool improvements [5].

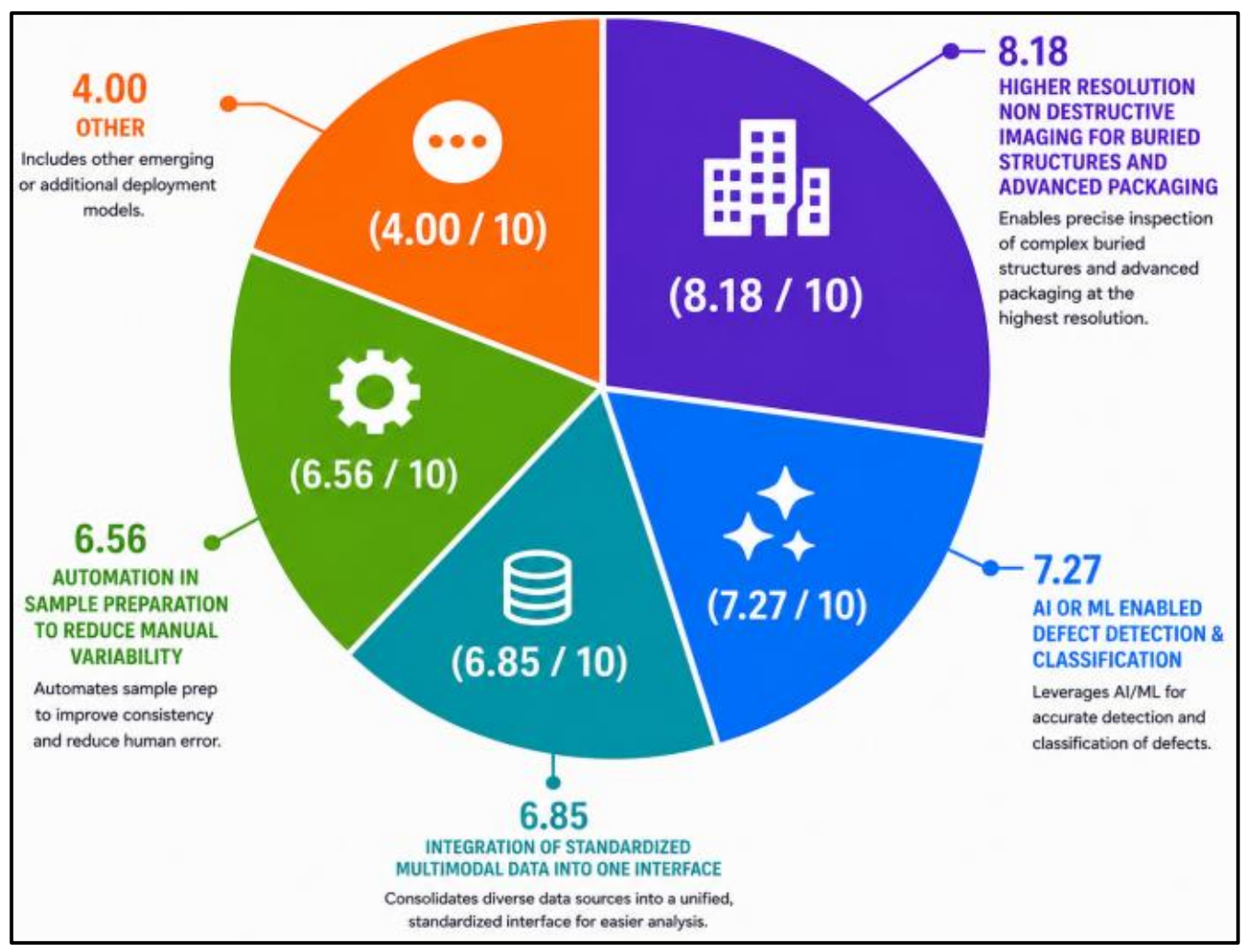

Fig. 4. Key improvement areas expected to accelerate failure analysis workflows

Fig. 4. Depicts the average importance ratings of emerging failure analysis technologies, highlighting the significance of advanced imaging, AI-assisted defect classification, data integration, and automated sample preparation in next-generation semiconductor packaging. This result supports a practical roadmap for future development. First, buried structures must become more visible with sufficient resolution & throughput. Second, the outputs of multiple modalities must become easier to align, compare, & interpret. Third, artificial intelligence can become useful only when the underlying data is standardized enough to support trustworthy training & validation. The survey therefore does not present AI as a replacement for expert failure analysis. Instead, it positions AI as an accelerator that becomes credible only after improvements in imaging quality, data quality, & workflow coherence [6].

## VIII. STANDARDIZATION & DATA INFRASTRUCTURE

One of the strongest consensus results in the survey concerns standardization. A clear 83% of respondents stated that the industry needs formalized data standardization frameworks for advanced packaging technologies such as fan out wafer level packaging [7], 2.5D interposers [8], 3D stacked dies [9], & hybrid bonding [10], while only 17% disagreed. The associated open-ended comments explain why. Participants asked for standardized interconnect & interface definitions, common failure analysis data formats & handoff procedures, stronger statistical reporting practices, benchmark datasets for AI & machine learning, standardized techniques for sample preparation, tool integration with EDA & test ecosystems, & end to end traceability metadata that captures electrical & thermomechanical intent.

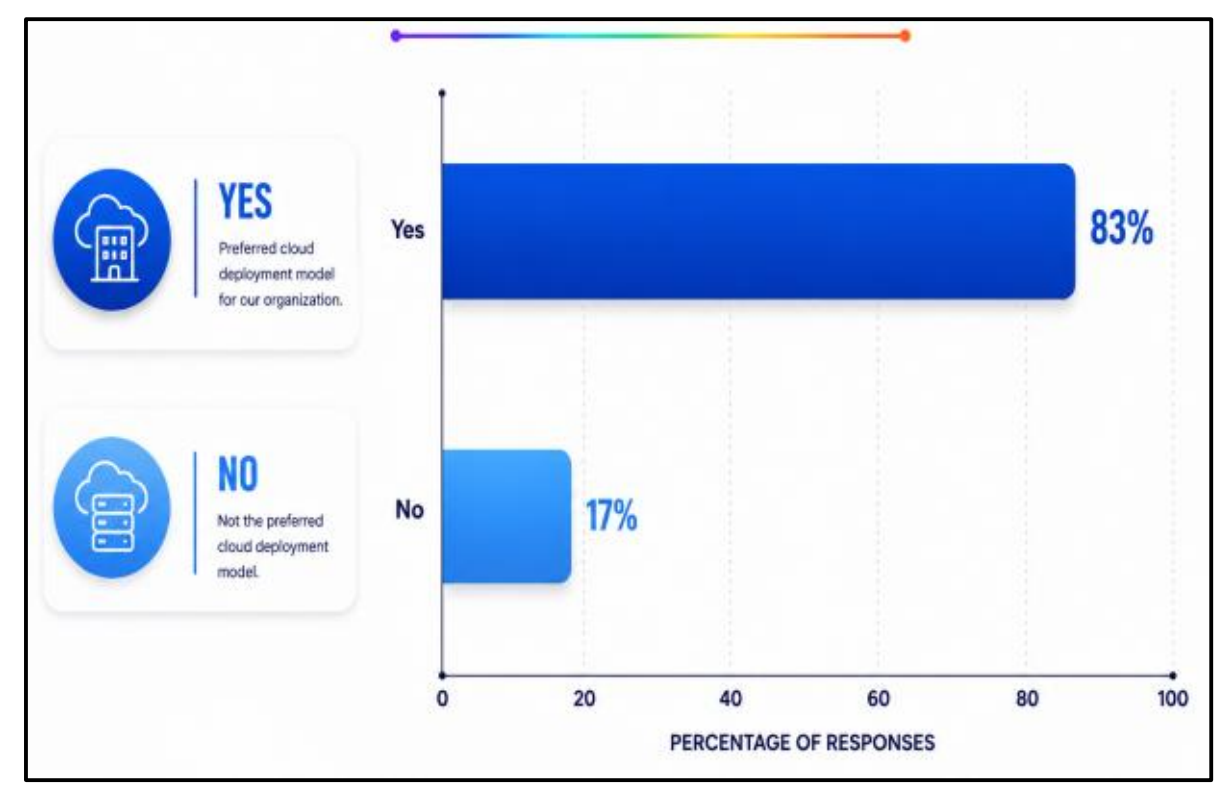

Fig. 5. Industry needs formalized data standardization frameworks in advanced packaging failure analysis

Fig. 5. survey responses indicating the perceived need for additional failure analysis tools and methodologies in advanced semiconductor technologies. A significant majority of respondents (83%) agreed that current approaches require further development to address the growing complexity of modern packaging and heterogeneous integration architectures. At the same time, the minority perspective is important. The respondents who opposed standardization did not reject rigor; instead, they warned that a single rigid framework may not fit the wide diversity of problems, manufacturers, & applications. They also noted that over standardization could suppress creativity & may be difficult to apply in proprietary environments. The implication is that the most effective path is likely a layered standardization model. Core metadata, reporting structures, & multimodal handoff conventions can be standardized, while detailed analysis strategies remain flexible enough to accommodate product specific constraints.

## IX. DISCUSSION

Taken together, the survey results support three larger conclusions about the trajectory of failure analysis. First, the center of gravity of failure analysis is moving from a predominantly device centric discipline toward package, interface, & system level diagnostics. This transition does not reduce the importance of device level methods. Rather, it indicates that future FA must increasingly address packaging complexity, heterogeneous integration, & cross domain interactions among electrical, thermal, mechanical, optical, & material domains. This shift is especially important for emerging technologies such as silicon photonics, co packaged optics, hybrid electronic photonic integration, backside power delivery, & three-dimensional integration, where failure mechanisms may no longer be confined to a single transistor, metal line, or die. In these systems, FA must account for optical loss, thermal drift, coupling efficiency, waveguide damage, interfacial delamination, contamination, stress induced misalignment, & package level degradation as interconnected contributors to performance loss.

Second, the hardest problems identified by the survey are coupled problems. Hybrid bonding, backside power delivery, silicon photonics, & three-dimensional integration all require simultaneous control of access, imaging, deprocessing fidelity, & interpretation. For example, in photonic systems, conventional electrical FA alone may be insufficient because failures can arise from changes in optical mode confinement, fiber to chip coupling, resonator detuning, heater instability, or mechanically induced waveguide deformation. Similarly, in hybrid bonded & vertically integrated systems, the relevant defect may be buried beneath multiple functional layers & may require a combination of nondestructive inspection, targeted cross sectioning, electrical characterization, & materials analysis. These examples suggest that future FA workflows must become more multimodal, physics aware, & application specific.

Third, data infrastructure is becoming nearly as important as instrumentation. Without standardized multimodal data structures, process traceability, & links between inspection results & design intent, even high-quality measurements can remain isolated & difficult to convert into reusable engineering knowledge. A particularly important direction is the closer integration of FA within line & near line process monitoring. Performing failure analysis only after complete fabrication, assembly, or package level qualification can be expensive, time consuming, & sometimes too late to identify the root cause with sufficient confidence. Future FA should therefore move closer to the manufacturing flow, where early indicators such as wafer warpage, bond interface nonuniformity, optical alignment drift, contamination signatures, dielectric damage, or process induced stress can be detected before additional value is added through subsequent process steps.

This in line direction also connects naturally to the development of digital twin frameworks for advanced electronics manufacturing & packaging. A digital twin could combine design files, process parameters, inspection data, metrology outputs, electrical test results, optical measurements, & physical FA observations into a unified representation of the device or package. Such a framework would allow FA results to be used not only for post failure diagnosis, but also for predictive learning, process correction, yield improvement, & reliability forecasting, making FA part of a closed-loop engineering system rather than a final-stage activity.

A particularly important finding is that the survey aligns physical bottlenecks with organizational priorities. The respondents are not merely asking for better microscopes or better algorithms in isolation. They are implicitly pointing toward an integrated workflow that begins with early localization & process aware monitoring, proceeds through minimally invasive imaging, transitions into targeted destructive analysis when necessary, & ends with a data framework capable of linking observations to design intent, process history, & system behavior. This is a more demanding model of failure analysis than the one traditionally associated with single die debug, but it is the model increasingly required by advanced packaging, heterogeneous integration, silicon photonics, & future system level electronics.

## X. CONCLUSION

This paper analyzed an industry informed survey of approximately one hundred participants regarding the present & future needs of failure analysis in advanced semiconductor technologies. The results show that heterogeneous integration, chiplet, & three-dimensional products now dominate the respondent landscape, & that package & heterogeneous integration failure analysis has become the highest rated strategic focus area. Hybrid bonding was identified as the hardest new architecture to analyze, while higher resolution non-destructive imaging was ranked as the most important accelerator, & standardized data frameworks as a broadly shared industry need.

The broader message is that failure analysis is no longer well described as a sequence of disconnected inspection steps. In advanced packaging & heterogeneous integration, it must be treated as a coordinated diagnostic system that connects materials, structures, interfaces, package mechanics, design context, & data infrastructure. Survey based studies such as this one are valuable because they expose where practitioners see the greatest friction in that system. The complete survey data, presented in Appendix A (Table I), supports transparency, replication, & benchmarking for future work. The present results indicate that the future of failure analysis will depend on simultaneous progress in non-destructive observability, artifact aware sample preparation, architecture specific diagnostic strategies, & flexible but formalized data integration frameworks.

## ACKNOWLEDGEMENT

The authors gratefully acknowledge the support of the Florida Semiconductor ENGINE program for facilitating this survey study. The support provided through the NSF Engine contributed to the development, coordination, & dissemination of the survey, enabling engagement with participants from relevant academic, industry, & research communities.

# APPENDIX A

**TABLE II.** ***Complete Industry Survey Results: Failure Analysis in Advanced Packaging & Heterogeneous Integration***

*The following table presents all quantitative & qualitative findings from the survey described in the main text. It is included to improve transparency, enable benchmarking, & support future comparative studies. Percentages reflect share of respondents selecting each option; importance ratings are on a 10-point scale.*

| Survey Question / Category | Response / Finding | Result / Statistic |
|---|---|---|
| ***A. Respondent Profile and Organizational Background*** | | |
| Organization type (select all that apply) | OSAT and packaging organizations | **22%** |
| | Fabless companies | **19%** |
| | Failure analysis laboratories | **16%** |
| | Systems and OEM organizations | **10%** |
| | EDA and tool vendors | **10%** |
| | Other (academic, modeling/simulation, prototyping, material suppliers) | **23%** |
| ***B. Supported Product Domains*** | | |
| Primary product domain(s) supported (select all that apply) | Heterogeneous integration, chiplet, and 3D products | **69%** |
| | Photonics and co-packaged optics | **41%** |
| | Logic at advanced nodes | **35%** |
| | Power products | **24%** |
| | MEMS and sensors | **20%** |
| | Memory | **16%** |
| | Other | **14%** |
| ***C. Importance of Failure Analysis Areas (rated on a 10-point scale)*** | | |
| Rate the importance of each FA domain for the coming years | Package and heterogeneous integration failure analysis | **7.92 / 10** |
| | System-level failure analysis | **7.56 / 10** |
| | Device-level fault isolation | **7.51 / 10** |
| | Device-level physical analysis | **7.03 / 10** |
| ***D. Sample Preparation and Imaging Limitations*** | | |
| Which type of analysis method most needs improvement? | Non-destructive analysis approaches | **72%** |
| | Destructive analysis approaches | **28%** |
| Open-ended requests for non-destructive imaging improvements | Higher resolution; greater penetration depth; faster throughput; improved 3D reconstruction; multimodal imaging; 3D fault localization without disturbing assembly; nano-void detection; high-resolution 3D X-ray; AI under limited data and trustworthiness constraints | **Multiple responses** |
| Open-ended requests for destructive analysis improvements | Cryogenic SEM for fragile materials; low-dose TEM; sub-angstrom metrology; ultra-thin lamella preparation | **Multiple responses** |
| Sample preparation challenges cited (open-ended) | Ultra-thin lamella without artifacts; accurate delayering in advanced BEOL/low-k; access to BSPDNs, TSVs, TGVs, hybrid bonded interfaces; handling large/warped/non-standard geometries | **Multiple responses** |
| ***E. Emerging Architecture Pain Points*** | | |
| Which new architecture is hardest to analyze today? | Hybrid bonding | **54%** |
| | Gate-all-around (GAA) devices | **24%** |
| | Backside power delivery networks (BSPDN) | **14%** |
| | Other | **8%** |
| Hybrid bonding: primary challenges (open-ended) | Detecting voids, gaps, and misalignment at the bonding interface | **Most cited** |
| | Cross-section preparation without inducing artifacts | **2nd most cited** |
| | Managing warped wafers / dies and impact on bonding yield | **3rd most cited** |
| GAA devices: dominant challenge | Structural damage during electrical probing / nanoprobing | **56%** |
| | Delayering without losing device integrity | **44%** |
| BSPDN: dominant challenge | Electrical probing / nanoprobing from the backside | **75%** |
| | Locating and isolating defects in buried power rails | **25%** |
| 3D heterogeneous integration: challenges cited (open-ended) | Heat dissipation; EMC; active embedded components; die-to-die fault isolation; CTE mismatch; hybrid bonding quality; power delivery; signal integrity; cross-die and cross-abstraction-level interactions | **Multiple responses** |
| ***F. What Would Most Accelerate Future Failure Analysis (rated on a 10-point scale)*** | | |
| Rate each proposed improvement area | Higher-resolution non-destructive imaging for buried structures / advanced packaging | **8.18 / 10** |
| | AI / ML-enabled defect detection and classification | **7.27 / 10** |
| | Integration of standardized multimodal data (EFA, PFA, X-ray, acoustic) | **6.85 / 10** |
| | Automation in sample preparation | **6.56 / 10** |
| ***G. Standardization and Data Infrastructure*** | | |
| Does the industry need formalized data standardization frameworks for advanced packaging FA? | Support for formalized frameworks | **83%** |
| | Concerns about rigidity, proprietary constraints, suppression of creativity | **17%** |
| Specific standardization requests (open-ended) | Standardized interconnect/interface definitions; common FA data formats and handoff procedures; stronger statistical reporting; benchmark datasets for AI/ML; standardized sample preparation techniques; tool integration with EDA and test ecosystems; end-to-end traceability metadata capturing electrical and thermomechanical intent. | **Multiple responses** |